\documentclass[final,5p,times,twocolumn]{elsarticle}

\usepackage{graphicx}
\usepackage{amssymb}
\usepackage{amsthm}
\usepackage{amsmath}
\usepackage{lineno}
\usepackage{subfigure}

\usepackage{upgreek}

\biboptions{sort&compress}

\usepackage{booktabs}

\journal{Journal of molecular liquids}

\begin{document}

\begin{frontmatter}


\title{The influence of an applied magnetic field on the self-assembly of magnetic nanogels}



\author[First]{Ivan S. Novikau$^\ast$}
\ead{ivan.novikau@univie.ac.at}
\author[Second,Third]{Pedro A. S{\'a}nchez}
\author[First,Third]{Sofia S. Kantorovich}

\address[First]{Faculty of Physics, University of Vienna, Boltzmanngasse 5, 1090 Vienna, Austria}
\address[Second]{Helmholtz-Zentrum Dresden-Rossendorf, Bautzner Landstrasse 400, 01328 Dresden, Germany}
\address[Third]{Ural Federal University, Lenin Av. 51, Ekaterinburg 620000, Russian Federation}

\begin{abstract}
Using Langevin dynamics simulations, we investigate the self-assembly of magnetic nanogels in the presence of applied magnetic fields of moderate strength. We find that even weak fields lead to drastic changes in the structure factors of both, the embedded magnetic nanoparticles and of whole nanogel particles. Nanogels assemble by uniting magnetic particle clusters forming inter-gel bridges. At zero field the average amount of such bridges for a pair of nanogels is close to one, whereas even for weak fields it fastly doubles. Rapid growth of cluster size at low values of the applied field is followed by a broad region of slow increase, caused by the mechanical constraints imposed the polymer matrix. The influence of the latter manifests itself in both, the slow growth of the magnetisation curve at intermediate fields and the slow decay of the total Zeeman energy.
\end{abstract}


\end{frontmatter}


\section{Introduction}

The name `microgel' was introduced more than six decades ago to describe soft colloidal particles, with characteristic sizes of up to several microns, made of a permanently crosslinked network of diluted polymers \cite{Baker1949,2007-aleman-pac,2011-fernandez-nieves-bk, 2017-hamzah-jpr}. Nowadays, the growing importance of nanoscale systems has introduced a practical distinction between microgels---with sizes ranging from 0.1 to 100~$\upmu$m---and nanogels---with sizes up to 100~nm---that is becoming widespread.

Micro- and nanogel particles are one of the most interesting approaches to create smart materials, as they can be made responsive to many different stimuli, including pH \cite{Hoare2004,Yin2008}, temperature \cite{Hoare2004,Mohanty2015,Backes2017a}, electromagnetic radiation \cite{Gorelikov2004}, ionic strength or electric fields \cite{Mohanty2015, Colla2018}. Their responses include large structural changes, typically swelling/collapse transitions, and a complex rheology \cite{2018-doukas-sm,2017-foglino-prl}. Such a plethora of responsive possibilities became available in recent years thanks to the rapid development of synthesis techniques \cite{2011-fernandez-nieves-bk, Bonham2014, 2016-mavila-chrv, 2017-hamzah-jpr}. The most common among such techniques are the ones based on the polymerization of diluted monomers in presence of crosslinking agents, either in homogeneous solutions or in emulsion droplets. Alternatively, network crosslinking can be achieved by photoinduced generation of radicals or by electrochemical methods \cite{2016-mavila-chrv, 2016-galia-eccm}.

The large variety of properties that micro- and nanogels may exhibit proves them to have a great potential for many technological and biomedical applications \cite{Das2006, 2011-thorne-cps, Hu2012, Bonham2014,Agrawal2018}. They serve to create responsive coatings \cite{Raquois1995,Saatweber1996}, chemical sensors and biosensing probes \cite{RETAMA2003,Guo2005,2014-sigolaeva-bmm,Aliberti2017}, as well as in tissue engineering \cite{Pepe2017}, manipulation and template-based synthesis of solid nanoparticles \cite{Bayliss2011, Zhang2004a} or water management and oil or pollutant recovery \cite{Son2016,Alhuraishawy2018,PU2019}. Nanogels have a particular relevance for drug delivery systems, as their structural changes can control the confinement and release of cargos whereas their reduced size allows them to cross biological barriers that are insurmountable obstacles for larger particles \cite{LOPEZ2005,2008-kwon-oh-pps,Vinogradov2010,2011-malmsten-ch,Sivakumaran2011,Pepe2017,Schimka2017}.

In difference with other responsive behaviours, that are mainly associated to the physical and chemical properties of the polymers forming the soft particle, a response to external magnetic fields can be settled for micro- and nanogels by embedding magnetic nanoparticles into the polymer network. This idea was introduced for the creation of macroscopic magnetic gels \cite{1995-shiga, 2000-zrinyi, 2010-reinicke, 2011-frickel, 2011-messing, 2013-ilg, 2013-xu, 2015-roeder} and soon adopted for the synthesis of microscopic gel particles \cite{2004-menager-pol,2015-backes-jpcb}. The use of magnetic fields as control stimuli allows to avoid possible side effects induced by changing parameters to which many soft matter substances---particularly biochemical compounds---are sensitive, such as pH or electric field. In most soft matter systems, only the magnetic nanoparticles introduced by design have a significant response to magnetic fields of reasonable strength.

The importance of micro- and nanogels stimulated the application of many different approaches, experimental and theoretical, to study their structural, mechanical, and rheological properties
\cite{2006-hoare-jpcb, 2018-backes-pol, 2019-witte-sm,Kobayashi2014,Ghavami2016,Kobayashi2016,Ahualli2017,Ghavami2017,Kobayashi2017,Gnan2017,Hofzumahaus2018, 2019-martin-molina-jml-rev, 2020-scotti-flow}. In particular, computer simulations on these systems have experienced an important development in very recent years with the adoption of a more realistic representation of the polymer network, moving away from the lattice or regular structures used in former models \cite{Gnan2017,Moreno2018,2019-rovigatti-sm-rev}. In this context, two years ago we presented the first coarse-grained simulation model of magnetic nanogels with a non regular internal structure \cite{Minina2018}. This model, that also can represent non magnetic gel particles \cite{2019-minina-jml}, is intended to reproduce qualitatively the structural features of nanogels obtained by electrochemically or photonically induced crosslinking of polymer precursors confined in emulsion nanodroplets \cite{2016-galia-eccm,2016-mavila-chrv,2019-minina-jml}.

In our preliminary studies, we employed the aforementioned  model to investigate the equilibrium structural properties of single magnetic nanogels \cite{Minina2018} and their suspensions \cite{2019-novikau-icmf} in absence of an applied external field. The latter work allowed us to understand the influence of nanogel concentration and the impact of interparticle magnetic interaction strength on the structural properties of the suspensions. However, to the best of our knowledge, the response of magnetic nanogels to applied magnetic fields---an essential property for any practical application---remains unexplored. With this manuscript we aim at filling this gap by elucidating the magnetic response of nanogel suspension to external fields.

The structure of the manuscript is the following. In Section \ref{sec:sim-m} we discuss the main ingredients of the simulation model and protocol. Results are presented in Section \ref{sec:rnd} and discussed in two main parts: in Section \ref{subsec:struct}, we calculate structure factors parallel and perpendicular to the field for both, embedded magnetic nanoparticles and whole nanogels; in that section we also perform cluster analysis. Next, in Section \ref{subsec:resp} magnetisation and magnetic energies as functions of the applied magnetic field are discussed. Finally, conclusions and outlook are provided in section \ref{sec:con}.

\section{Simulation Approach}\label{sec:sim-m}
Our magnetic nanogel (MNG) model is based on a classical bead-spring representation of the polymer chains \cite{Kremer1990}, with the addition of a fraction of embedded magnetic particles and crosslinks \cite{Minina2018, 2019-minina-jml}. The setup of each MNG is performed using the following procedure. Polymer precursors are represented as chains of spherical beads with dimensionless unit diameter, $\sigma =1$,  and mass, $m=1$. A steric repulsion between them is given by a shifted and truncated Lennard-Jones (LJ) potential, or Weeks-Chandler-Andersen (WCA) potential \cite{1971-weeks}:
\begin{equation}
U_{W C A}(r)=\left\{\begin{array}{ll}{4\left[r^{-12}-r^{-6}\right]+1,} & {r \leqslant 2^{1/6}} \\ {0,} & {r > 2^{1/6}}\end{array}\right. ,
\label{eq:wca}
\end{equation}
where $r$ is the centre-to-centre distance between the interacting beads measured in $\sigma$. Here, the depth of the LJ potential well is set to unity. This defines the energy scale in our system.  Additionally, thermal energy, $k_BT$ is also set to unity, resulting in all our interaction energies being dimensionless and measured in units of $k_BT$. The centres of neighbouring beads along the polymer chains are connected by means of FENE springs, forming the polymer backbones. FENE potential has the form:
\begin{equation}
U_{F E N E}(r)=-\frac{1}{2} \epsilon_{f} r_f^2 \ln \left[1-\left(\frac{r}{r_{f}}\right)^{2}\right],
\label{eq:fene}
\end{equation}
where $r_{f}=1.5$ is the maximum bond extension and $\epsilon_{f}=22.5$ is the dimensionless bond rigidity parameter. MNGs are created by placing initially $N_{p}=6$ linear polymer chains, with $L=100$ beads each, inside a spherical confinement. The volume  fraction of beads inside the spherical confinement is $\phi_p \approx 0.1$. Once the polymer ensemble is equilibrated, interchain crosslinks are randomly established according to a minimum interparticle distance criteria, up to a fraction of crosslinks $\phi_{\mathrm{links}}=0.17$. Each added crosslink is an elastic spring that links the centres of the newly joined particle pair, described by harmonic potential:
\begin{equation}
U_{h}(r)=-\frac{1}{2} Kr^{2},
\label{eq:harm}
\end{equation}
with a dimensionless elastic constant $K=10$. After crosslinking is performed, the spherical confinement is removed. For more details on the crosslinking protocol, see \cite{2019-minina-jml}. For the sake of simplicity, magnetic particles are introduced by placing permanent magnetic dipoles, $\vec \mu$, in the centres of randomly chosen beads, up to 60 magnetic particles per MNG. Such  beads represent single domain ferromagnetic nanoparticles, that interact by means of the magnetic dipole-dipole pair potential:
\begin{equation}
U_{d d}\left(\vec{r}_{i j}\right)=\frac{\left(\vec{\mu}_{i} \cdot \vec{\mu}_{j}\right)}{r^{3}}-\frac{3\left(\vec{\mu}_{i} \cdot \vec{r}_{i j}\right)\left(\vec{\mu}_{j} \cdot \vec{r}_{i j}\right)}{r^{5}},
\label{eq:dipdip}
\end{equation}
\begin{figure}[h!]
	\centering
		\includegraphics[width=0.46\textwidth]{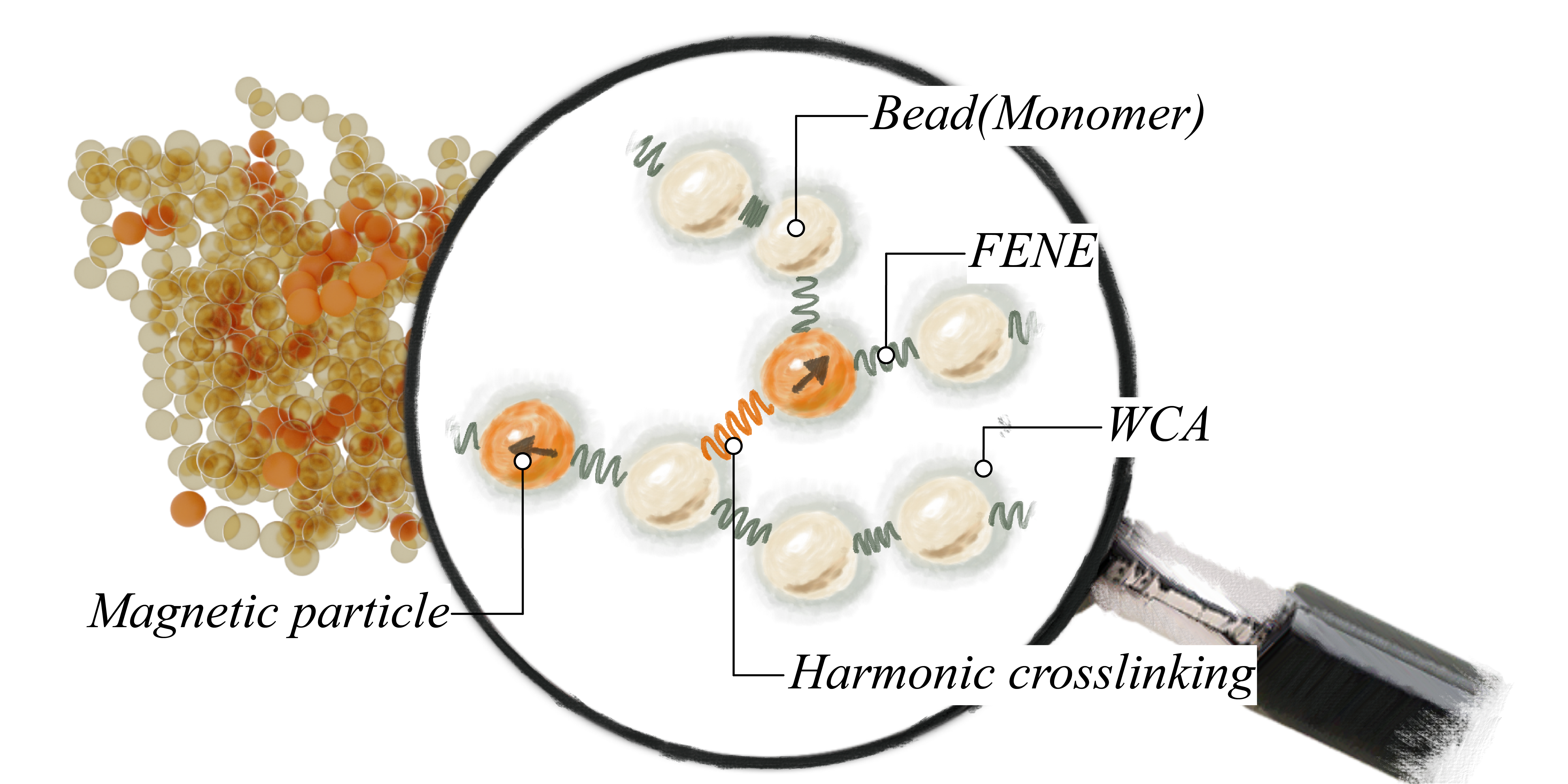}
	\caption{The sketch reflects the inner MNG bead-spring structure. The MNG itself is shown in the background at left side. Arrows inside particles represent the magnetic dipoles.}
	\label{fig:Model_beads}
\end{figure}
where $\vec \mu_i$, $\vec \mu_j$ are the respective dipole moments of the interacting particles and $\vec r_{ij}$ is the vector between their centres. 
 
In addition to interacting with each other, the dipoles are affected by an external constant uniform magnetic field, $\vec H$, directed along the $y$-axis. In simulations, $H$ is dimensionless, and defined as a so-called Langevin parameter $\alpha = \mu H/k_BT$, for unit magnetic moment, $\mu=1$. Potential of a single dipole under $\vec H$ is represented by the Zeeman energy:
\begin{equation}
U_{\mathrm{Zeeman}}=-\vec{\mu} \cdot \vec{H}.
\label{eq:Zeeman}
\end{equation}

In our dimensionless system the standard dipolar coupling parameter, $\lambda$, which is the ratio of the maximum strength of the dipole-dipole interaction to the thermal energy, can simply be defined as $\lambda=\mu^2$. In this study we take $\lambda=6$. With this value, in zero external magnetic field, the clusters of MNGs due to magnetic interparticle interactions just begin to form \cite{2019-novikau-icmf}. The dimensionless field strength was chosen to span  the range $|\vec{H}|=H \in [0,1.3]$. 

A characteristic snapshot with the zoomed-in sketch of the MNG model can be found in Fig.~\ref{fig:Model_beads}.

A suspension is created by randomly placing $N=100$ MNGs, derived from the process described above, in a cubic box with volume $V = 96^3$ and periodic boundary conditions. Therefore, the volume fraction of MNGs was equal to 0.1; whereas the volume fraction of magnetic material was below one per cent. It should be noted that each individual MNG is equilibrated prior to be placed in the suspension. In total, 10 different equilibrium configurations with distinct magnetic particle and cross-links intrinsic distributions are used to create the suspensions. This helps to avoid any dependence of the self-assembly on the individual MNG topology. Molecular dynamics simulations in the NVT ensemble with a Langevin thermostat and dimensionless thermal energy ($k_BT=1$, as mentioned above) is performed using the simulation software package {ESPResSo} \cite{2019-weik}. The dipolar P$^3$M algorithm is employed to calculate long-range magnetic interactions \cite{2008-cerda-jcp}. Using a fixed time step $\delta t=0.01$, the system is first equilibrated by making $2 \times 10^{7}$ integration steps. Over the next $9 \times 10^{6}$ integration steps, measurements are collected. Each set of parameters is sampled using eight separate runs, so that not only time-, but also ensemble averaging is performed, in order to improve statistics.

\section{Results and Discussions}\label{sec:rnd}
\subsection{Impact of an applied field on the structural properties}\label{subsec:struct}

One of the standard methods to investigate the internal structure of soft matter is to perform small angle light, X-ray or neutron scattering \cite{1983-kotlarchyk-analysis,1997-pedersen-analysis,2002-fernandez-structural,2003-gazeau-static,2004-stieger-small,2004-stieger-are,2008-grillo-small,2004-avdeev-atructural,2019-ninarello-modeling}. In case of a spatially isotropic system, the scattering intensity can be converted into a centre-centre structure factor (SF) $S$:
\begin{equation}
S(\vec q)=\frac{1}{N}\sum_{j=1}^N\sum_{k=1}^N e^{-i\vec q \cdot (\vec r_j - \vec r_k)},
\label{eq:sf}
\end{equation}
here $N$ is the number of scattering centres in the system ($N=100$ in case of calculating it for nanogel centres of mass, and $N=6000$ in case of SF of MNPs), $\vec q$ is the wave vector, and $\vec r_i$ is the coordinate of the $i$-th centre. If, as in our case, an external field is applied along $y$-axis, the scattering intensity becomes anisotropic, and it is convenient to define structure factors parallel and perpendicular to $\vec{H}$. To calculate the structure factor in the plane perpendicular to the applied magnetic field, $S(\vec q_{\bot})$, it is enough to take $y$ component of the vector $\vec q$ equal to zero and use the formula \eqref{eq:sf}. Similar, in order to calculate the structural factor along the field, $S(\vec q_{\parallel})$, the $x$ and $z$ components of $\vec q$ must be zero.

\begin{figure}[h!]
	\centering
	\subfigure[]{\label{fig:sf-magn-xy}\includegraphics[width=0.23\textwidth]{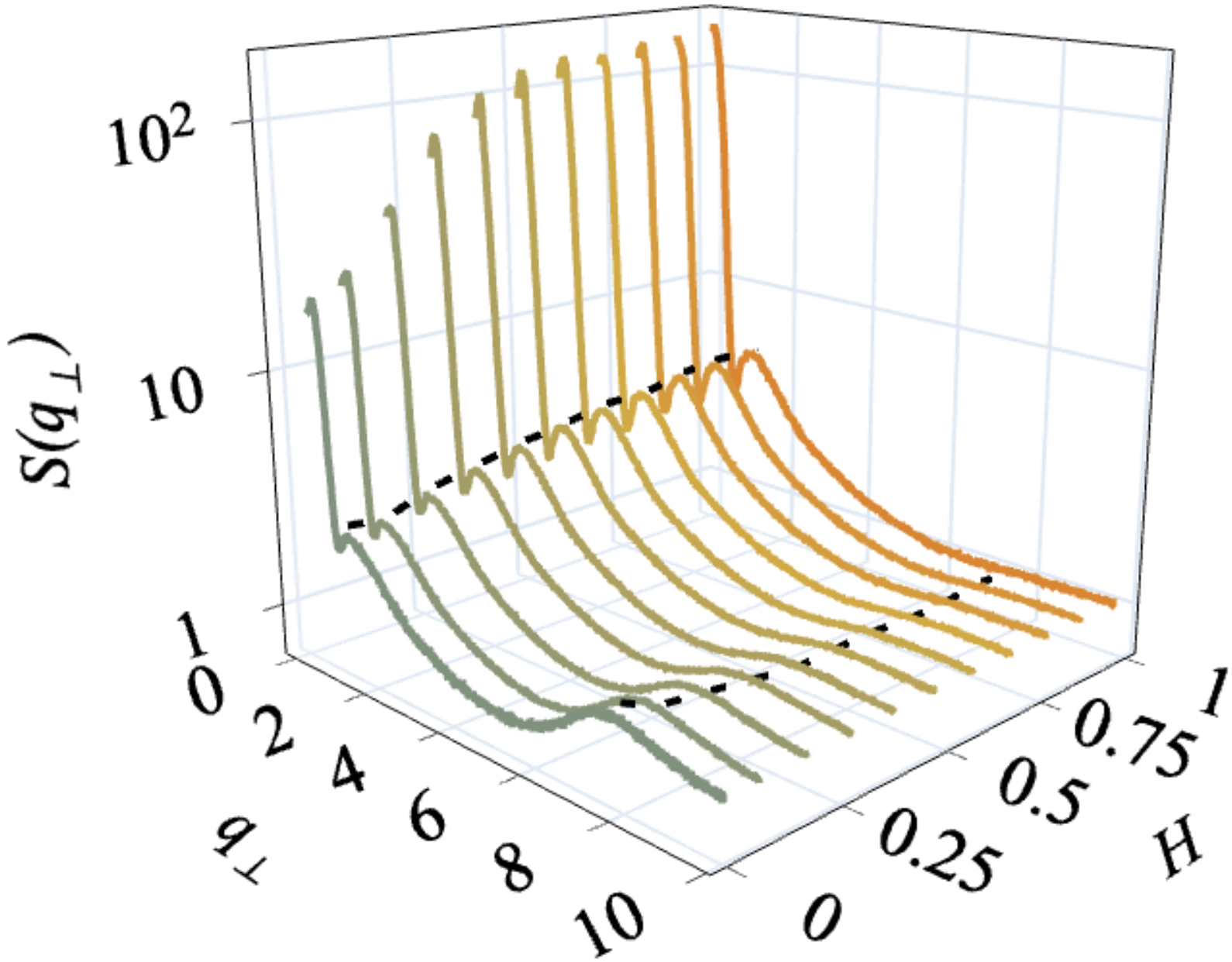}}
	\subfigure[]{\label{fig:sf-magn-z}\includegraphics[width=0.23\textwidth]{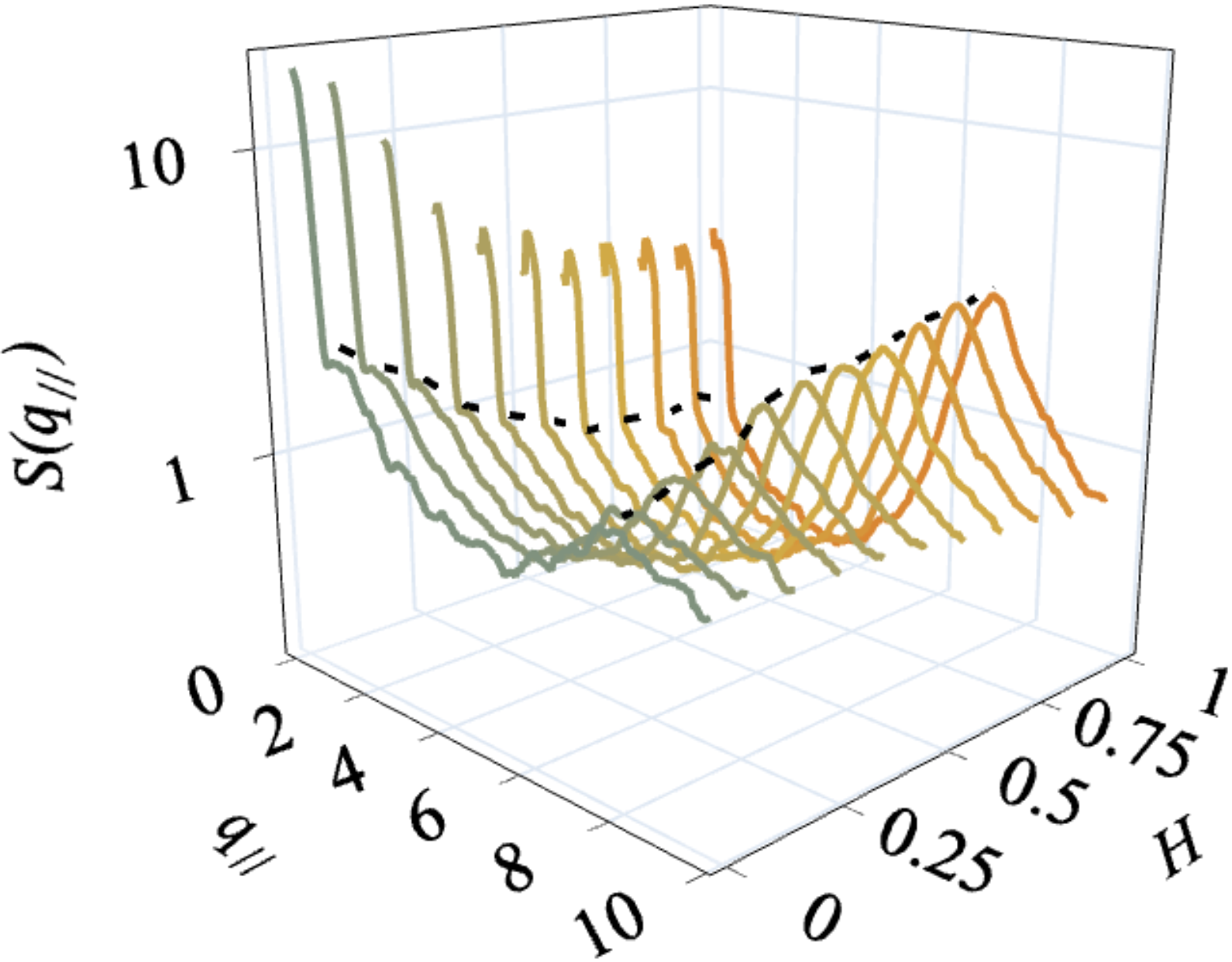}}
    \subfigure[]{\label{fig:sf-cm-xy}\includegraphics[width=0.23\textwidth]{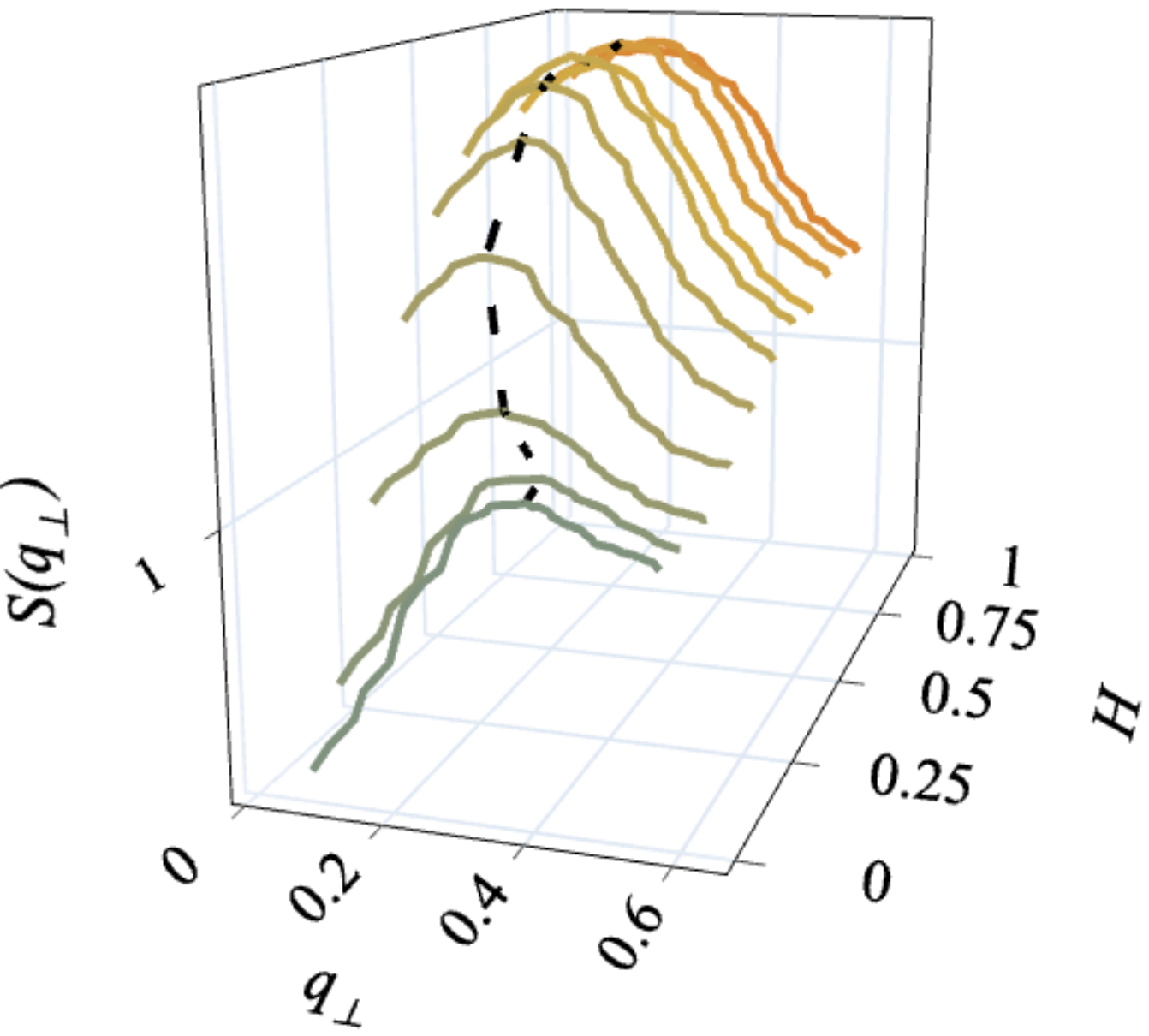}}
	\subfigure[]{\label{fig:sf-cm-z}\includegraphics[width=0.23\textwidth]{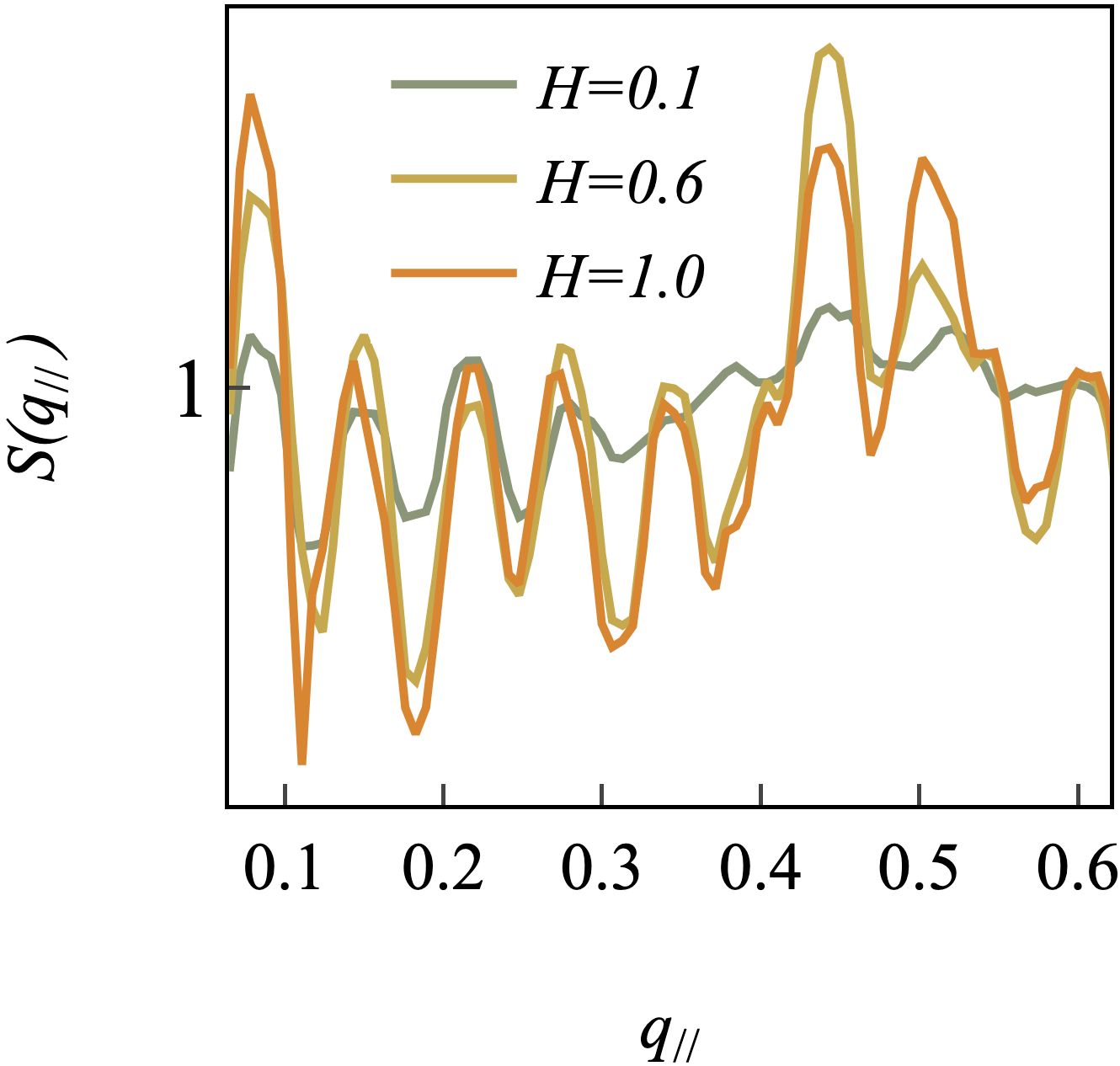}}
	\caption{Structure factors $S(\vec q_{\bot})$ and $S(\vec q_{\parallel})$ calculated perpendicular (a) and (c) and parallel (b) and (d) to an external applied magnetic field $H$, $\vec{H}= \left(0,H,0\right)$. (a) and (b) are calculated for magnetic particles. (c) and (d) are obtained for nanogel centres of mass.}
	\label{fig:sf-field}
\end{figure}

In Fig.~\ref{fig:sf-field} we plot $S(\vec q_{\bot})$ and $S(\vec q_{\parallel})$ as a function of $|\vec{q_{\bot}}| =q_{\bot}$ and $|\vec{q_{\parallel}}| =q_{\parallel}$ respectively, for magnetic particles in the upper row and the same observables for the nanogel centres of mass in the lower row. 
The field $H=0.25$ corresponds to a Zeeman energy almost two times smaller than $kT$ and more than 10 times smaller than the interaction of coaligned dipoles of two magnetic particles in close contact. This rather weak field, however, turns out to be sufficient to cause significant changes in the structuring of nanogel magnetic particles perpendicular to the field, as it can be seen in In Fig.~\ref{fig:sf-magn-xy}. The peak at $q_{\bot}\sim 7$, corresponding to a close contact of two magnetic particles, vanishes with growing intensity of applied field, manifesting the reorientation or reconfiguration of the magnetic particle clusters in the plane perpendicular to the field direction. At the same time, in Fig.~\ref{fig:sf-magn-z}, where $S(\vec q_{\parallel})$ is plotted, the same peak rapidly grows up to $H=0.5$. For higher fields, the height of the peak keeps growing in Fig.~\ref{fig:sf-magn-z} and decreasing in Fig.~\ref{fig:sf-magn-xy}, but with a noticeably lower rate. The peak at $q_{\bot}\sim 1$, that in real space corresponds to the distance of approximately 6-7 diameters of magnetic particles, grows in Fig.~\ref{fig:sf-magn-xy} and shifts to the right with increasing field, whereas in Fig.~\ref{fig:sf-magn-z} it basically disappears as $H$ grows. This behaviour reflects the field-induced change of the inner structure of the nanogels. If no field is applied, magnetic particle clusters are distributed rather homogeneously inside a nanogel and have a characteristic distance between them defined by the gel size and magnetic particle concentration. Once $H>0$, then magnetic particle chains grow longer and tend to align with the field as much as the elastic matrix around them allows, thus, leading to a slight decrease of the characteristic distance between them in the direction perpendicular to the field, and basically spanning through nanogels in the direction parallel to $H$. This behaviour of magnetic particles indicates a tendency of nanogels to aggregate under the influence of an applied magnetic field.

This conjecture is fully confirmed by Figs.~\ref{fig:sf-cm-xy} and \ref{fig:sf-cm-z}, where we plot structure factors calculated for nanogels centres of mass. Here, the peak at $q\sim 0.4$, corresponds to the contact of two nanogels in real space. This peak grows and shifts to the left in Fig.~\ref{fig:sf-cm-xy}, while in Fig.~\ref{fig:sf-cm-z} it splits into repeated subpeaks. In other words, we observe a liquid-like structure with increasing intergel distance in the direction perpendicular to $H$, and a clear nanogel chain formation in the direction of the field. As in the case of magnetic nanoparticles discussed above, in Figs.~\ref{fig:sf-cm-xy} and \ref{fig:sf-cm-z} even a small field leads to significant changes. 

\begin{figure}[h!]
	\centering
	\subfigure[]{\label{fig:clusters}\includegraphics[width=0.23\textwidth]{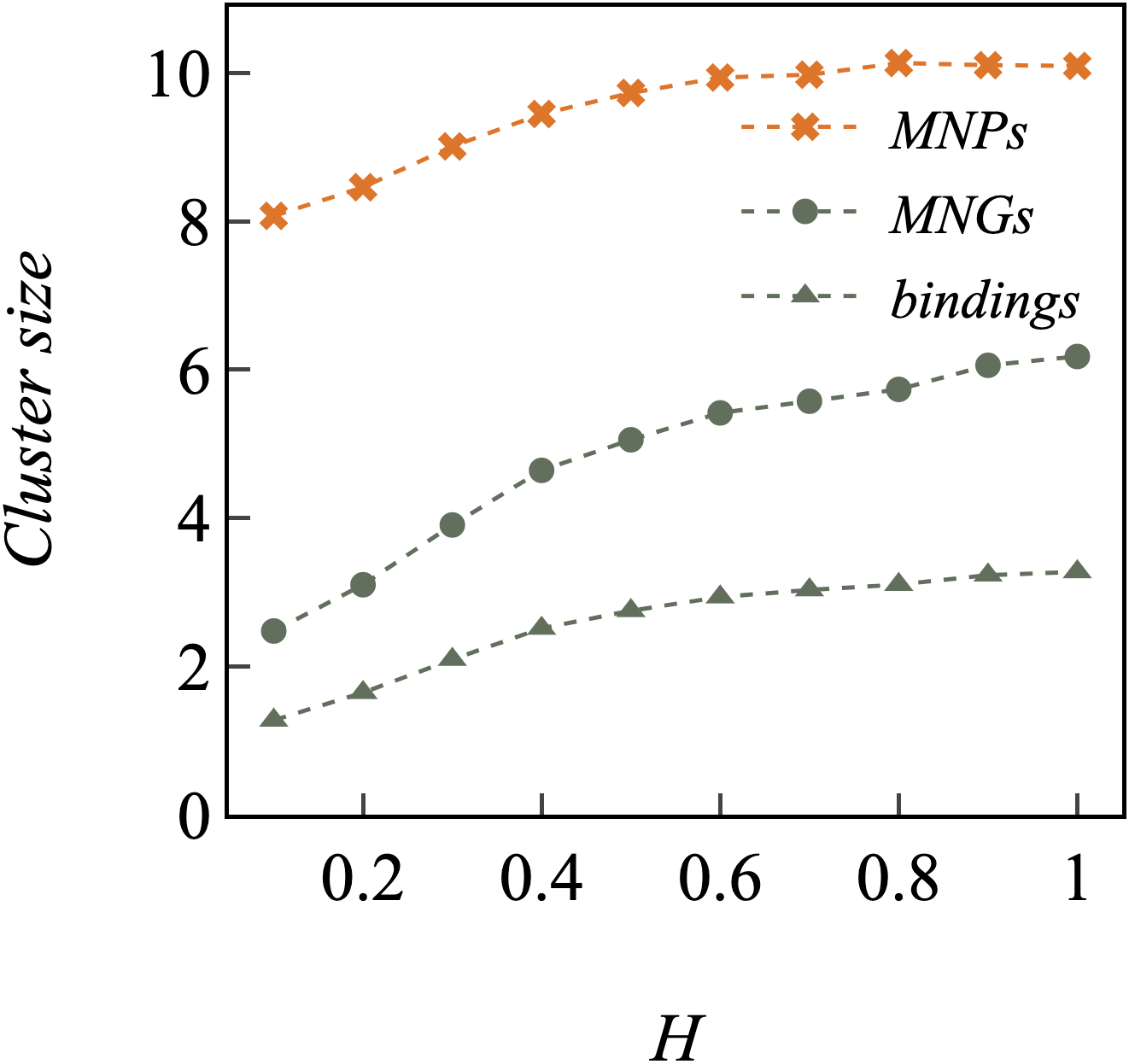}}
	\subfigure[]{\label{fig:clusters-pic}\includegraphics[width=0.23\textwidth]{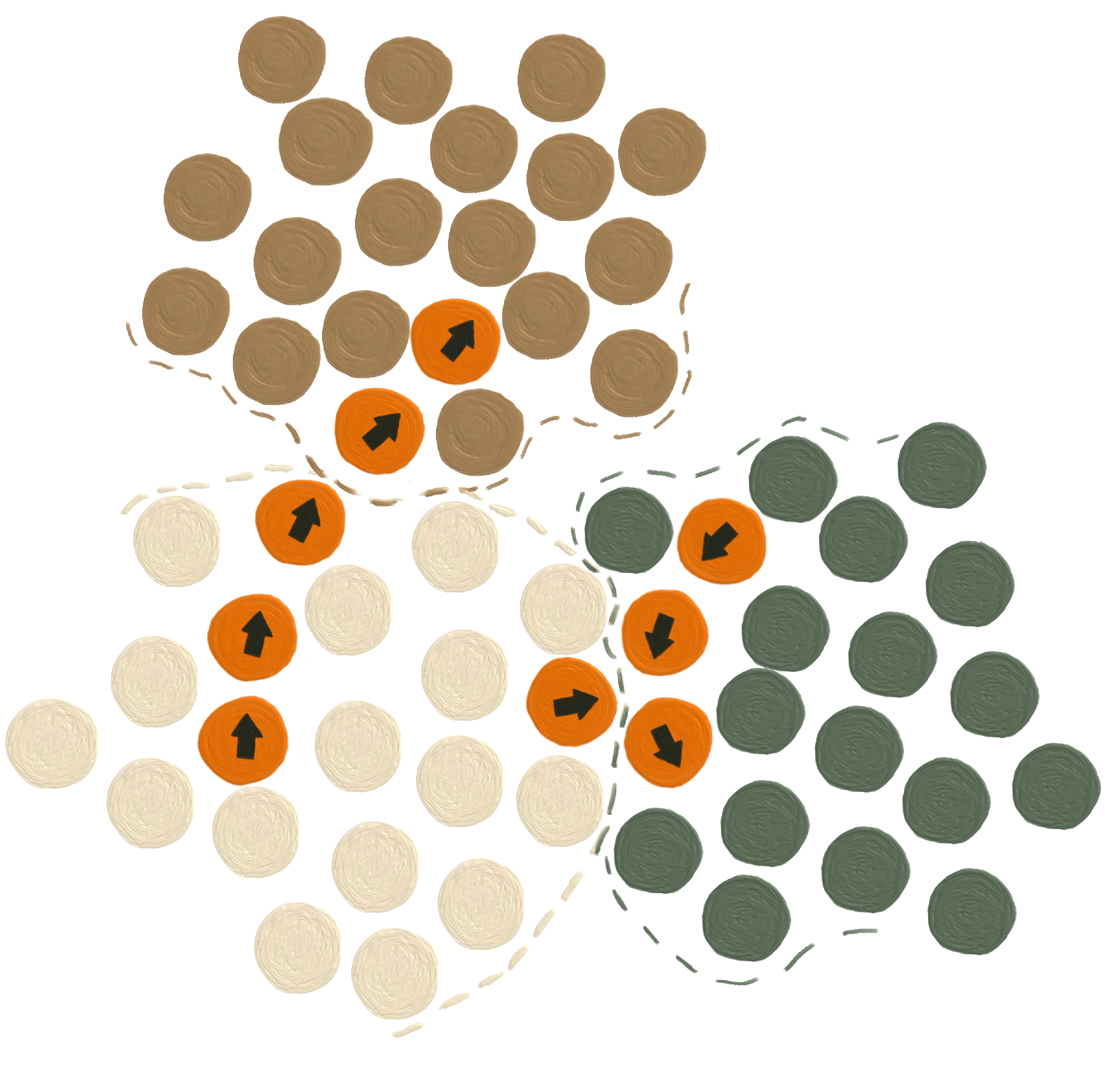}}
	\caption{(a) Average cluster size as a function of $H$. Upper curve shows the average number of magnetic particles in their clusters; middle curve -- the average number of nanogels in their clusters; the lowest curve shows the average amount of bridges per a dimer of nanogels. (b) Schematic representation of three nanogels. Two nanogels on the left are forming a dimer, as the common cluster has more than four particles (orange), out of which at least two belong to each of the nanogels. The left low and the right nanogels are not forming a cluster as the condition to have at least two particles of a bridge in each nanogel is violated in the left low nanogel.}
	\label{fig:rdf-conc-l}
\end{figure}

In order to obtain a more quantitative picture of the chain formation, in Fig.~\ref{fig:clusters} we plot three different curves characterising average cluster size. The upper curve shows the average size of clusters formed by magnetic nanoparticles only, in which two particles are considered to be connected according to the distance-energy criterion \cite{2009-pyanzina}, {\it i.e.}, if their interaction energy is negative and the distance between their centres is below 1.5.  It can be seen that the average number of magnetic particles in the cluster rapidly grows with an applied field up to $H\sim 0.5$. After that, the growth slows down significantly. The same characteristic behaviour is exhibited by the middle curve, that shows the average size of clusters of whole nanogels: it grows almost by a factor of two when field changes from zero to $H=0.5$, whereas the further increase of the field up to $H=1$ results in only 20\% enhancement of nanogel aggregation. We did not investigate here the behaviour for larger values of $H$, as the cluster size could exceed half of the simulation box and lead to the finite-size induced percolation. However, for the range of parameters described above, the system size is appropriate. In our work, two nanogels are considered to be connected if they share a cluster of magnetic particles and each nanogel contains at least two particles of such cluster. According to this, in Fig.~\ref{fig:clusters-pic}, the upper and the lower left nanogels are forming a cluster, whereas the right low nanogel is not a member of it, as only one magnetic particle of a shared cluster belongs to the left low nanogel. In case two nanogels are bonded, we call the shared cluster an intergel bridge. The average number of intergel bridges per nanogel as a function of the applied field is described by the lowest curve in Fig.~\ref{fig:clusters-pic}. One can see that, with growing field, each nanogel has on average two bridges. The behaviour of this curve, similar to the two upper ones, changes once the value of $H$ exceeds 0.5. 

Summarising the investigation of field-induced structural transformations in the suspension of magnetic nanogels, one can say that even if the coupling between magnetic moments and an external field is of the same order of thermal energy,  extensive self-assembly, both on the level of magnetic nanoparticles and nanogels, takes place as suggested by the anisotropy of the scattering properties and confirmed by direct cluster analysis. Compared to field-induced aggregation in ferrofluids, this strong impact of an applied weak field on the assembly of magnetic nanogels may be surprising. It is related to the fact that, locally, within each magnetic nanogel the concentration of nanoparticles is higher than in a homogeneous ferrofluid with the same overall volume fraction of magnetic material ($\sim$ 0.4 per cent). This induces a local increase of magnetic correlations. In absence of applied field, these correlations lead to chain formation inside the nanogels, however, the mainly random orientation of these chains leads only to moderate intergel correlations. Once a field is applied, even if it is weak, chain aggregates tend to reorient due to their high susceptibility \cite{2004-mendelev}, providing nanogels with relatively large magnetic moments oriented along the field, giving rise to long-range pronounced intergel interactions. The latter results in the field-induced self-assembly discussed in this Section. Why at a certain, still rather weak field, the character of the field-dependent structural transformations changes and the cluster size growth notably slows down? The reason is the presence of the elastic constraints that limit the mobility of magnetic particles, thus preventing the formation of additional intergel bridges, or larger clusters inside the gels.

In order to get a better understanding of the magnetic response of the suspension,  in the next subsection we study magnetisation and dependence of the potential energy of the system on the applied magnetic field. 
\begin{figure*}[h!]
	\centering
		\subfigure[]{\label{fig:magn-c}\includegraphics[width=0.69\textwidth]{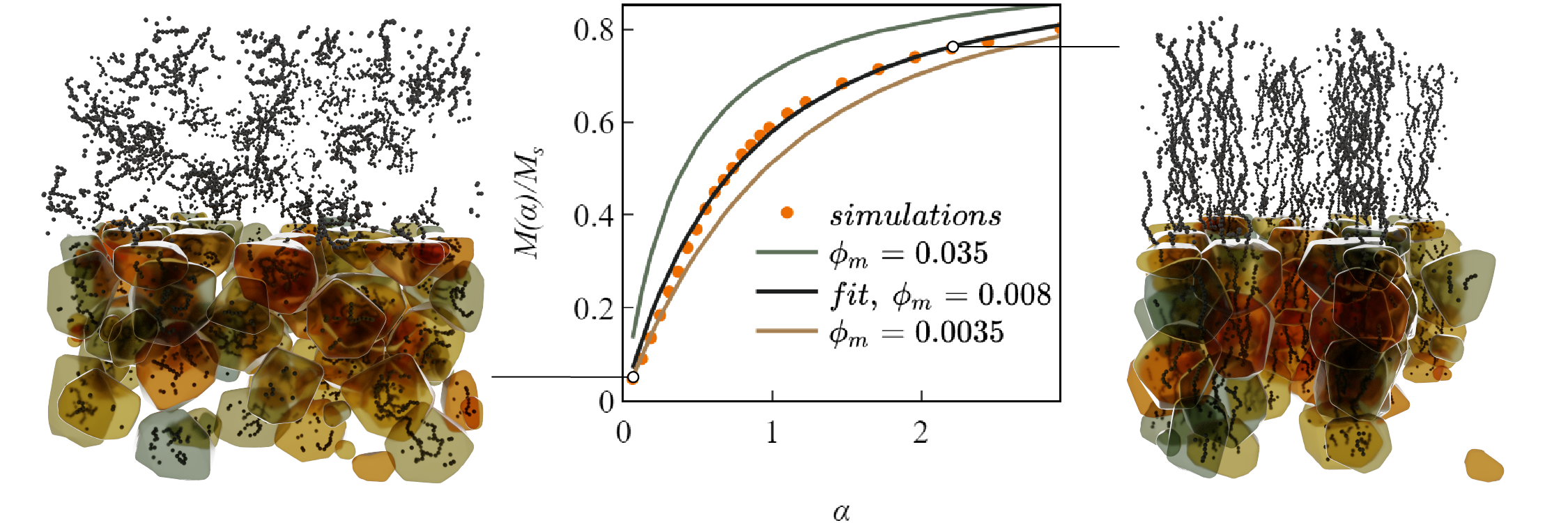}}
		\subfigure[]{\label{fig:clust-s-t}\includegraphics[width=0.23\textwidth]{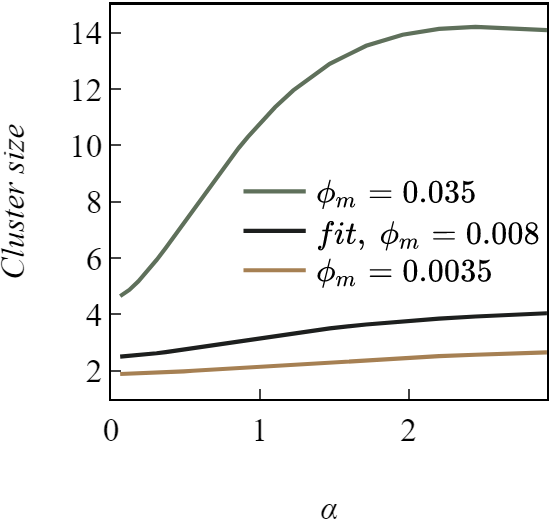}}
	\caption{(a) Magnetisation curves and selected simulation snapshots. In the latter, magnetic particles are represented as solid black spheres, while only the contours of the polymer matrices are represented as translucent surfaces. The latter are absent in the upper part of the snapshots to ease the visualisation of the arrangements of magnetic particles. (b) Average cluster size as a function of $H$. The curve in the middle shows the average number of magnetic particles for the fit $\phi_m = 0.008$; upper curve -- $\phi_m = 0.035$; the lowest curve -- $\phi_m = 0.0035$. }
	\label{fig:magnetisation}
\end{figure*}
\subsection{Magnetic response} \label{subsec:resp}

Magnetisation can be calculated as an average projection of the sum of all magnetic moments on the field axis. In Fig.~\ref{fig:magn-c} we plot the magnetisation of the suspension of magnetic nanogels as a function of  Langevin parameter $\alpha=\mu H /(kT) = \sqrt{\lambda} H$ with symbols. Two simulation snapshots, corresponding to selected values of $\alpha$ are also shown. The ordering under the influence of an applied field can be easily seen on the right snapshot. The magnetisation curve of the nanogel suspension alone cannot tell much about the system, as from the first glance it looks like a typical Langevin-like law \cite{langevin1905a,langevin1905b}. However, it is well known that for $\lambda =6$, for a system of noncrosslinked magnetic particles, one has to use the model of magnetisation that explicitly takes into account the formation of chains \cite{2004-mendelev}. This model, based on the density functional theory, predicts a rapid growth of the initial susceptibility with $\lambda$, attributed to the formation of highly correlated and responsive to weak magnetic fields ferroparticle chains and was shown to accurately describe magnetisation curves in both real and computer experiments with low concentrated systems of magnetic particles \cite{2006-ivanov,klokkenburg08a}. As we saw in the previous section, the chains of magnetic particles indeed reorient in very weak fields inside the nanogels and form intergel bridges. On the other hand, it is known that polymer matrix hinders the magnetisation of magnetic gels \cite{2015-weeber}.

In order to understand the competing mechanisms involved, in Fig.~\ref{fig:magn-c} we plot, together with the simulation data, three theoretical curves. The lowest curve was calculated using the volume fraction of magnetic particles $\phi_m = 6000 \pi/(6V) = 0.0035$, namely the actual content of magnetic material in the simulation box. As it can clearly be seen, the effects of locally higher fraction of magnetic particles inside nanogels results in a higher magnetisation than such theoretical expectation, apart from the very low field region: $\alpha<0.2$, $H<0.2/\sqrt{6} \sim 0.08$. If we consider that the magnetic particles are encapsulated inside nanogels, we can estimate $\phi_m = 60/(8R_g^3)$, where $R_g$ is the average gyration radius of a nanogel, $R_g\sim 6$. This estimate results in $\phi_m = 0.035$ and corresponds to the upper curve in Fig.~\ref{fig:magn-c}. This approach clearly overestimates simulation magnetisation in the whole region of external fields. An attempt to fit the data using the volume fraction as a fitting parameter results in $\phi_m=0.008$. This curve (in the middle) fits rather well the data, but has qualitatively different behaviour. Analogous behaviour of the magnetisation fit can be obtained for $\phi_m =0.035$, but using dipolar strength as a fitting constant, and obtaining and $\lambda = 5.2$. However, the same qualitative differences remain: chain model overestimates the initial slope ($H<0.2$) and underestimates slightly the intermediate range of $\alpha$, $0.2<H<0.8$; whereas for larger values of the applied field, the chain model magnetisation becomes again higher than that of the suspension of nanogels. 

Looking carefully at Fig.~\ref{fig:clust-s-t}, we can safely attribute these qualitative differences to the effects of the polymer matrix. Here, average cluster sizes, as predicted by the chain model \cite{2004-mendelev}, are plotted for $\phi_m =0.035, \ 0.008$ and 0.0035, being respectively distributed from top to bottom. It turns out that for $\phi_m = 0.008$, the average cluster size is almost twice smaller than we find in simulation (see, Fig.~\ref{fig:clusters}) even though the magnetisation curve appears to be very close to the simulation data. If we consider magnetic particles constrained in the volume of a nanogel, but free to move, $\phi_m = 0.035$ (upper curve in Fig.~\ref{fig:clust-s-t}) we see that the maximum value is around 14, while the upper curve in Fig.~\ref{fig:clusters} goes up to ten. Even more striking is the low-field region difference: for unconstrained particles the chain length is about four particles, but inside the nanogels this value is almost twice higher.

\begin{figure}[h!]
	\centering
	\subfigure[]{\label{fig:dip}\includegraphics[width=0.23\textwidth]{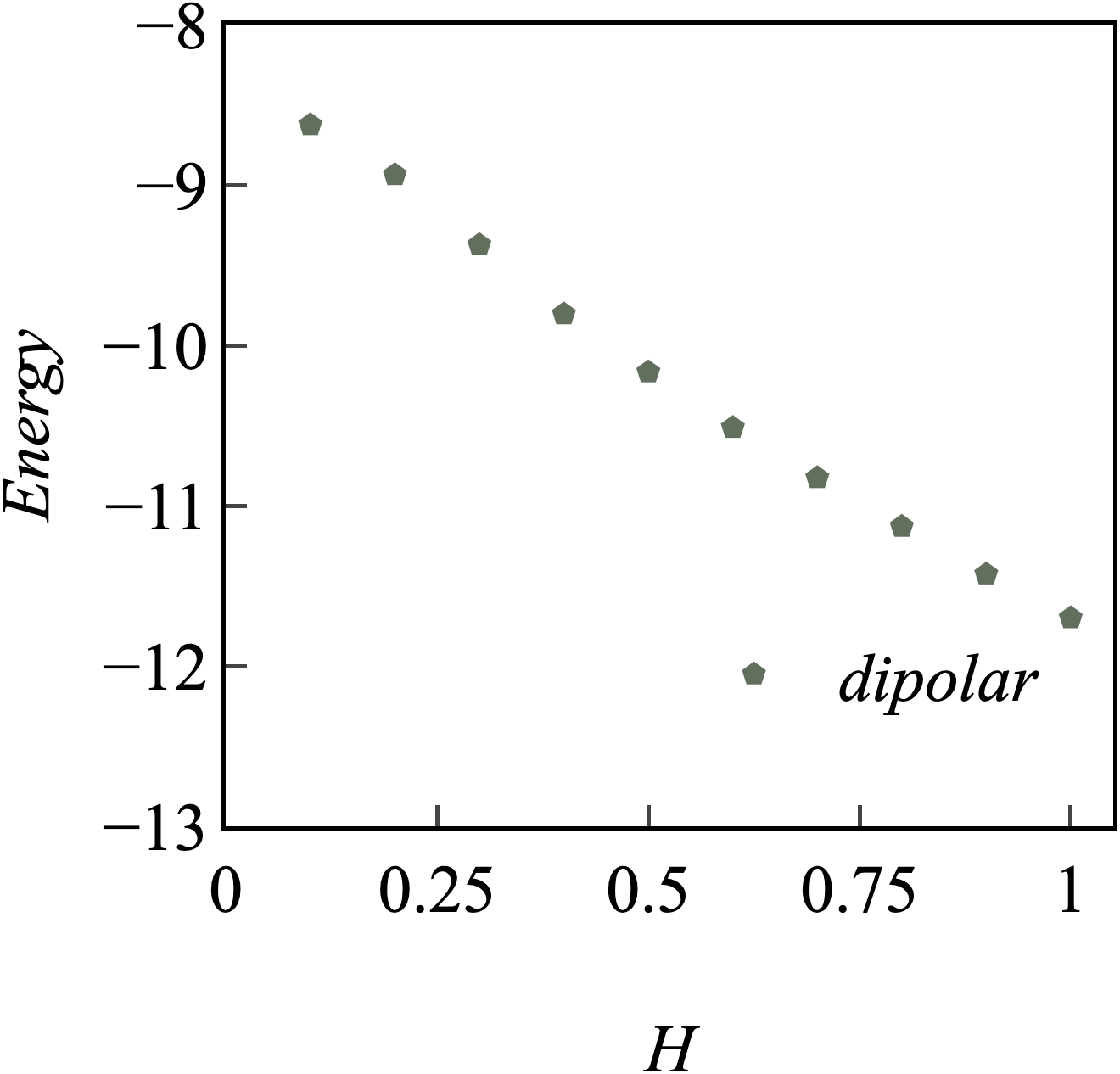}}
	\subfigure[]{\label{fig:zeeman}\includegraphics[width=0.23\textwidth]{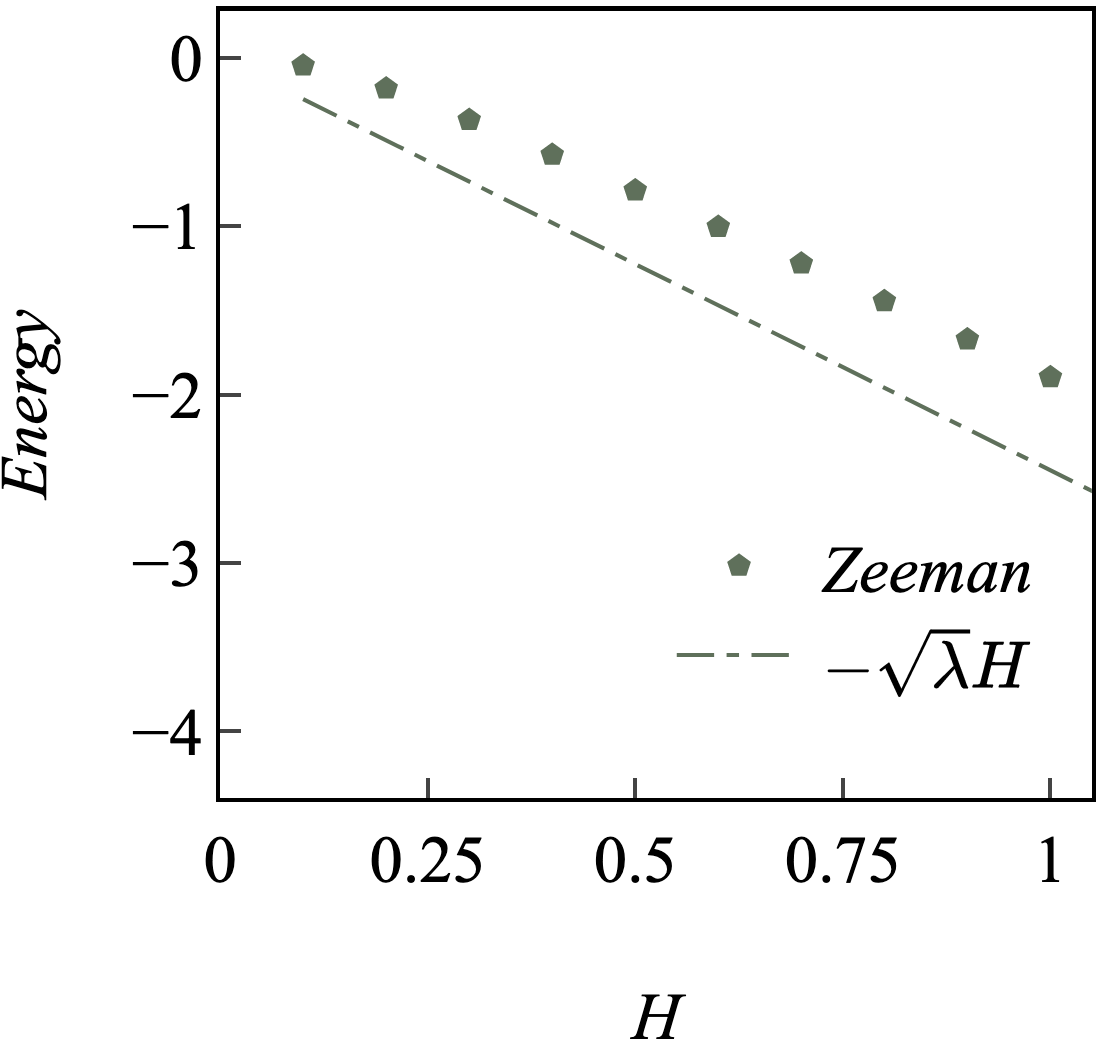}}
	\caption{Energy per particle calculated as a function of $H$. (a) Dipolar energy. (b) Zeeman energy; here, the dashed line corresponds to the full alignment of the dipole along the field $-\sqrt{\lambda}$H.}
	\label{fig:energies}
\end{figure}

Thus, the effect of the nanogel matrix and nanogel neighbours are competing: on the one hand, neighbouring nanogels allow to build intergel bridges and thus form longer chains, optimising dipolar interactions; on the other hand, matrix constraints inside individual nanogels set a limit to the chain length growth and restrict the response leading to lower overall magnetisation. This explanation can be confirmed by Fig.~\ref{fig:energies}, where dipolar and Zeeman energies per particle, obtained from simulations, are plotted against the dimensionless magnetic field $H$.
In Fig.~\ref{fig:dip} one can see that dipolar energy per particle rapidly decreases with $H$. Instead, the Zeeman energy, plotted in Fig.~\ref{fig:zeeman} is well above its limiting behaviour $-\sqrt{\lambda}H$.

Summarising this part of the investigation, one can say that the suspension of magnetic nanogels responds to an applied magnetic field in a qualitatively different manner than an analogous ferrofluid with unconstrained particles. The effects of the elastic constraints within the volume of each nanogel is twofold: they set the length limitations for the magnetic particles chain growth and, at the same time, hinders the net magnetisation along the field. The presence of neighbouring nanogels allows to partially reduce the frustration related to the chain growth by allowing to form longer intergel bridges, but still unconstrained particles would have formed larger aggregates under the same field.

\section{Conclusion}\label{sec:con}
We employed molecular dynamics computer simulations in order to investigate the influence of an applied magnetic field on the self-assembly of magnetic nanogels containing single domain ferromagnetic nanoparticles. We found that even weak fields lead to a very pronounced nanogel assembly in chains, aligned with the field direction. The reason for such a dramatic field influence is that magnetic nanoparticles are already highly correlated within the nanogels, so that an elastic-magnetic balance is achieved; while chains formed at zero field are randomly oriented, even a small field can reorient them as their susceptibility is very high. As soon as the chains reorient, they provide nanogels with a relatively large total magnetic moment and give rise to nanogel assembly. Moreover, the formation of intergel magnetic bridges becomes more advantageous as it allows to minimise dipolar energy forming longer chains without the need of large deformations of the polymer matrix.

As for magnetic response, it is found to be higher than for a ferrofluid with the same overall density of magnetic material; at the same time, magnetisation curve is found to be lower than that of a ferrofluid with the volume fraction of magnetic nanoparticles encapsulated inside each nanogel. In general, the behaviour of the magnetisation for the system of magnetic nanogels reveals qualitative differences from any ferrofluid, and the same is found for the average cluster size of magnetic nanoparticles, which for nanogels is found to be significantly larger at low fields, but smaller if the value of $H$ grows. Even though the clusters are larger at small fields, overall magnetic response is clearly hindered by the polymer matrix. As a result, the magnetisation is found to slowly reach the saturation, as well as the Zeeman energy decay with the field does not become linear for the fields considered here.

In a future work we plan to investigate the effects of higher magnetic fields. This is a rather costly task, as the system dimensions should be large enough to avoid finite size effects. Moreover, the rheology of these suspensions will be addressed in detail.

\section*{Acknowledgments}
This research has been supported by the Russian Science Foundation Grant 19-72-00209. Authors are grateful to A. O. Ivanov for valuable discussions concerning structure factors and to E. S. Pyanzina for providing the code for calculation of chain partition functions in an applied magnetic field. The work was also supported by the FWF START-Projekt Y 627-N27. Simulations were performed in the Vienna Scientific Cluster (VSC3).

\end{document}